\documentclass[aps,reprint,twocolumn,superscriptaddress,citeautoscript,showpacs,amsmath,amssymb,floatfix,prb]{revtex4}
\usepackage{graphicx}
\usepackage{amsmath}
\usepackage{amssymb}
\usepackage{mathtools}
\usepackage{textcomp,gensymb} %For degree symbol
\usepackage{bm} %For bold math (Greek) symbols or bold italic in math mode
\usepackage{float}
\usepackage{setspace}
\usepackage{dcolumn}
\usepackage{array}

\usepackage{changepage}
\usepackage{tikz}
\usepackage{contour}
\usepackage[left]{lineno}
\usepackage{blindtext}
\usepackage{setspace}
\usepackage[colorlinks=true,citecolor=blue,linkcolor=blue,pdftex,hypertexnames=false,urlcolor=blue,linktocpage=true]{hyperref}
\definecolor{ashgray}{rgb}{0.7,0.75,0.71}
\definecolor{mspringgreen}{rgb}{0, 0.8, 0.1}
\definecolor{auburn}{rgb}{0.43, 0.21, 0.1}
\definecolor{ao(english)}{rgb}{0.0, 0.5, 0.0}
\definecolor{afw}{rgb}{0.95, 0.95, 0.96}
\definecolor{magnolia}{rgb}{0.97, 0.96, 1.0}
\definecolor{wsmk}{rgb}{0.96, 0.96, 0.96}

\newcommand{\pst}{Pb$_{0.5}$Sn$_{0.5}$Te}
\newcommand{\psit}{(Pb$_{0.5}$Sn$_{0.5}$)$_{1-x}$In$_x$Te}

\newcolumntype{M}[1]{>{\centering\arraybackslash}m{#1}}

\usepackage{dcolumn}% Align table columns on decimal point

\newcolumntype{d}[1]{D{.}{.}{#1}}

\def\pbsnte{Pb$_{1-y}$Sn$_y$Te}

\begin{document}
%\title{Exploring the anomalous features in the lattice dynamics of \psit}
\title{Electron-phonon coupling and superconductivity in \\ the doped topological-crystalline insulator (Pb$_{0.5}$Sn$_{0.5}$)$_{1-x}$In$_x$Te}
\author{A. Sapkota}
\author{Y. Li}
\affiliation {Condensed Matter Physics and Material Science Department, Brookhaven National Laboratory, Upton, New York 11973, USA}
\author{B. L. Winn}
\author{A. Podlesnyak}
\affiliation{Neutron Scattering Division, Oak Ridge National Laboratory, Oak Ridge, Tennessee 37831, USA}
%\author{I. A. Zaliznyak}
%\affiliation {Condensed Matter Physics and Material Science Department, Brookhaven National Laboratory, Upton, New York 11973, USA}
\author{Guangyong Xu}
\affiliation{NIST Center for Neutron Research, National Institute of Standards and Technology, Gaithersburg, Maryland 20899, USA}
\author{Zhijun Xu}
\affiliation{NIST Center for Neutron Research, National Institute of Standards and Technology, Gaithersburg, Maryland 20899, USA}
\affiliation{Department of Materials Science and Engineering, University of Maryland, College Park, Maryland 20742, USA}
\author{Kejing~Ran}
\affiliation{School of Physical Science and Technology, ShanghaiTech University, Shanghai 201210, China}
\author{Tong Chen}
\affiliation{National Laboratory of Solid State Microstructures and Department of Physics, Nanjing University, Nanjing 210093, China}
\author{Jian Sun}
\author{Jinsheng Wen}
\affiliation{National Laboratory of Solid State Microstructures and Department of Physics, Nanjing University, Nanjing 210093, China}
\affiliation{Collaborative Innovation Center of Advanced Microstructures, Nanjing University, Nanjing 210093, China}
\author{Lihua Wu}
\author{Jihui Yang}
\affiliation{Materials Science \&\ Engineering Department, University of Washington, Seattle, WA 98195-2120, USA}
\author{Qiang Li}
\author{G. D. Gu}
\author{J. M. Tranquada}
\affiliation {Condensed Matter Physics and Material Science Department, Brookhaven National Laboratory, Upton, New York 11973, USA}

\date{\today}

\begin{abstract}
We present a neutron scattering study of phonons in single crystals of \psit\ with $x=0$ (metallic, but nonsuperconducting) and $x=0.2$ (nonmetallic normal state, but superconducting).  We map the phonon dispersions (more completely for $x=0$) and find general consistency with theoretical calculations, except for the transverse and longitudinal optical (TO and LO) modes at the Brillouin zone center. At low temperature, both modes are strongly damped but sit at a finite energy ($\sim4$~meV in both samples), shifting to higher energy at room temperature. These modes are soft due to a proximate structural instability driven by the sensitivity of Pb-Te and Sn-Te $p$-orbital hybridization to off-center displacements of the metal atoms. The impact of the soft optical modes on the low-energy acoustic modes is inferred from the low thermal conductivity, especially at low temperature.  Given that the strongest electron-phonon coupling is predicted for the LO mode, which should be similar for both studied compositions, it is intriguing that only the In-doped crystal is superconducting. In addition, we observe elastic diffuse (Huang) scattering that is qualitatively explained by the difference in Pb-Te and Sn-Te bond lengths within the lattice of randomly distributed Pb and Sn sites.  We also confirm the presence of anomalous diffuse low-energy atomic vibrations that we speculatively attribute to local fluctuations of individual Pb atoms between off-center sites.
\end{abstract}

\maketitle

\section{Introduction \label{Intro}}

The system \pbsnte\ has been of recent interest because of the theoretical prediction \cite{Hsieh_2012} and experimental confirmation \cite{Tanaka_2012,Xu_2012,Tanaka_2013,Ando_2015} that this material is a topological crystalline insulator \cite{Fu_2011} (for $y\gtrsim0.3$ \cite{Varjas_2020}). Furthermore, partial substitution of indium induces superconductivity, with a transition temperature, $T_c$, as high as 4.7~K \cite{Parfenev_2001,Zhong_2017}, while the normal state, from which the superconductivity develops, is nonmetallic, with a large resistivity \cite{Zhong_2014}.  

Calculations of phonons and electron-phonon coupling using density-functional perturbation theory \cite{Baroni_2001} yield an electron-phonon coupling constant that is large and sufficient to explain the superconducting transition temperature, $T_c$, in In-doped \pbsnte\ \cite{Ran_2018} and SnTe \cite{Nomoto_2020}, assuming that the assumptions behind the Allen-Dynes modification of the McMillan formula \cite{Allen_1975} are satisfied. There are reasons to question that for In-doped \pbsnte: 1) the strongest electron-phonon coupling involves the LO mode, which has the potential to be relatively soft at the zone center; 2) the non-metallic normal state \cite{Zhong_2014,Ran_2018} is inconsistent with the assumption of a Fermi liquid; 3) the value of 1.4 calculated for the electron-phonon coupling constant is near the limit of validity of Migdal-Eliashberg theory \cite{Esterlis_2018,Esterlis_2019A}.  %In fact, past studies suggest that this material should be near a structural instability \cite{Daughton_1978}.  
Indeed, while the calculated phonon density of states (PDOS) describes most of the experimental features for \psit, the observation of anomalous low-frequency modes \cite{Ran_2018} provided the motivation for the present investigation.

%While the lattice instability involves the same hybridized states that determine the topological character of this material, scanning tunneling microscopy (STM) measurements indicate an $s$-wave gap, consistent with a conventional superconducting state \cite{Du_2015}; nevertheless, a calculation for In-doped SnTe indicates that including spin-orbit coupling in the analysis leads to a higher $T_c$ \cite{Nomoto_2020}.  Another feature of interest is the role of the In dopants, and their potential to skip valence between $+1$ and $+3$, as when doped into SnTe \cite{Zhang_2018}.  Related behavior of Tl doped into PbTe, and its association with induced superconductivity, has been of interest for quite some time \cite{Matsushita_2005,Giraldo-Gallo_2018} 

In this paper, we apply inelastic neutron scattering (INS) to single-crystal samples of \psit\ with $x=0$ and 0.2 and determine their lattice dynamics.  With a more complete set of measurements on the $x=0$ crystal, we find that its normal modes are reasonably consistent with the phonons calculated by density-functional perturbation theory \cite{Ran_2018}.  An interesting difference is that, while the calculation was done for an insulating compound which can have a large LO-TO splitting, our sample is weakly metallic (probably due to vacancies), so that the LO and TO energies must be equal at $\Gamma$.  We find this energy is $\sim4$~meV and that the modes are strongly damped.  The $x=0.2$ crystal shows a similarly soft TO mode.

Beyond the normal modes, we demonstrate that there are no extra dispersive modes with zone-boundary energies of $<3$~meV.  Nevertheless, we do confirm evidence for diffuse excitations at energies below $\sim2$~meV, consistent with the previous measurements on powder samples \cite{Ran_2018}.  These excitations do not exhibit a conventional temperature dependence, having an intensity that appears to remain constant with temperature.  We propose that these excitations are due to off-center Pb ions independently fluctuating between different equivalent positions.

In the elastic channel, we observe Huang diffuse scattering.  This is qualitatively explained by the differences between the actual Pb-Te and Sn-Te bond lengths and the average bond-length of the lattice.  We also present a measurement of the thermal conductivity, $\kappa(T)$ for the $x=0$ crystal.  It is unusually low, especially at low temperature, where the LO and TO zone-center modes are at the minimum energy.  From the small value of $\kappa$ at low $T$, we infer that interactions of the LO and TO modes with the low-energy acoustic modes causes significant damping.

The rest of this paper is organized as follows.  In the following section, we discuss some of the background issues in a bit more detail.  Experimental details are given in Sec.~III.  The results are presented in Sec.~IV, where we begin with the conventional aspects of the phonon dispersions, consider the temperature and doping dependence of the soft zone-center TO modes, present the thermal conductivity, and analyze the lowest-energy zone-boundary dispersive modes.  We then present results for both elastic and inelastic diffuse scattering.  In Sec.~V, we discuss the possible causes of the low-energy diffuse scattering, and then consider some of the open questions regarding the nature of the superconductivity in this system.  A brief conclusion is presented in Sec.~VI.

\section{Background Issues in (P\lowercase{b}$_{1-y}$S\lowercase{n}$_{y}$)$_{1-x}$I\lowercase{n}$_x$T\lowercase{e}}

\subsection{Basic electronic structure}

If we consider a typical compound that adopts the rocksalt structure, such as an alkali halide, the electronic structure is fairly simple.  The alkali metal ion will transfer its one outer $s$ electron to the halogen ion, filling the outer $p$ shell of the latter.  This results in a large insulating gap of $\sim10$ eV, reflecting the lack of significant hybridization between these states.

In SnTe and PbTe, the situation is different.  The Te atom can accept two electrons, whereas the Sn and Pb each have two $s$ and two $p$ electrons in their outer shells.  If the Sn (Pb) were to give up all of its outer electrons, it would have a valence of $4+$, which would clearly be incompatible with the maximum charge of $2-$ on Te.

The DFT calculations and analysis of Waghmare {\it et al.} \cite{Waghmare_2003} provide a helpful picture of what actually happens.  The Sn $5s$ (Pb $6s$) states hybridize with the Te $5p$ states.  Because the binding energy of the $s$ state is relatively high, both the bonding and antibonding states are almost completely filled.  In contrast, the hybridization of Sn (Pb) $p$ states with Te $5p$ results in a filled bonding band and empty antibonding band; there is a small band gap between these states, where one would expect the chemical potential to lie. 

\subsection{Topological character}

In PbTe, the states at the top of the valence band are dominated by Te $5p$ character, while Pb $6p$ states dominate the bottom of the conduction band.  In SnTe, however, the states above the band gap, which occurs at the $L$ points, have Te $5p$ character.  As a result of this inversion of the electronic structure, SnTe is a topological crystalline insulator (TCI) \cite{Hsieh_2012,Fu_2011}.  The topological character derives from the mirror symmetry about a (110) plane.  When one dopes with Pb, the symmetry is no longer exact; nevertheless, calculations show that the topological character survives in the presence of disorder as long as the symmetry survives on average \cite{Ando_2015,Lusakowski_2018,Varjas_2020}.  Studies of \pbsnte\ with angle-resolved photoemission spectroscopy indicate that the transition from trivial insulator to TCI occurs near $y\sim0.3$ \cite{Xu_2012,Tanaka_2013,Yan_2014}.

It is natural to wonder whether the superconducting state in \psit\ could be topological.  If the superconducting state is time-reversal invariant, then one would expect the pairing to exhibit odd parity \cite{Michaeli_2012}.  As already mentioned, STM measurements on superconducting (Pb$_{0.5}$Sn$_{0.5}$)$_{0.7}$In$_{0.3}$Te indicate a gap with $s$-wave symmetry \cite{Du_2015}.  On the other hand, there is a recent report of Josephson-junction measurements on SnTe nanowires indicating an $s\pm is'$ gap, associated with time-reversal-symmetry breaking \cite{Trimble_2019}, so this issue might not yet be settled.  

\subsection{In doping}

The atomic orbital energy for In $5s$ electrons is a couple of electron volts higher than that for Sn $5s$.    As a result, when In is substituted into SnTe, the antibonding band with Te $5p$ is expected to sit near the chemical potential \cite{Tan_2016,Tan_2018}.  If these states are filled (empty), the In should act as a $+1$ ($+3$) ion. This is similar to the valence-skipping behavior of Tl dopants in PbTe, which has been of interest for quite some time \cite{Matsushita_2005,Geballe_2015,Giraldo-Gallo_2018}, 
%as reviewed by Geballe {\it et al.} \cite{Geballe_2015}, 
and it is part of a pattern common to other superconductors, such as BaBiO$_3$, as discussed by Varma \cite{Varma_1988}.   Our empirical observations suggest that the story is slightly more complicated: one has to consider whether or not these states are delocalized; a gradual crossover from localized to delocalized doped electron with increasing In concentration was indicated by transport measurements on In-doped SnTe \cite{Zhang_2018}.

\subsection{Structural instability}

\begin{figure}[t]
	\centering
	\includegraphics[width=0.8\columnwidth]{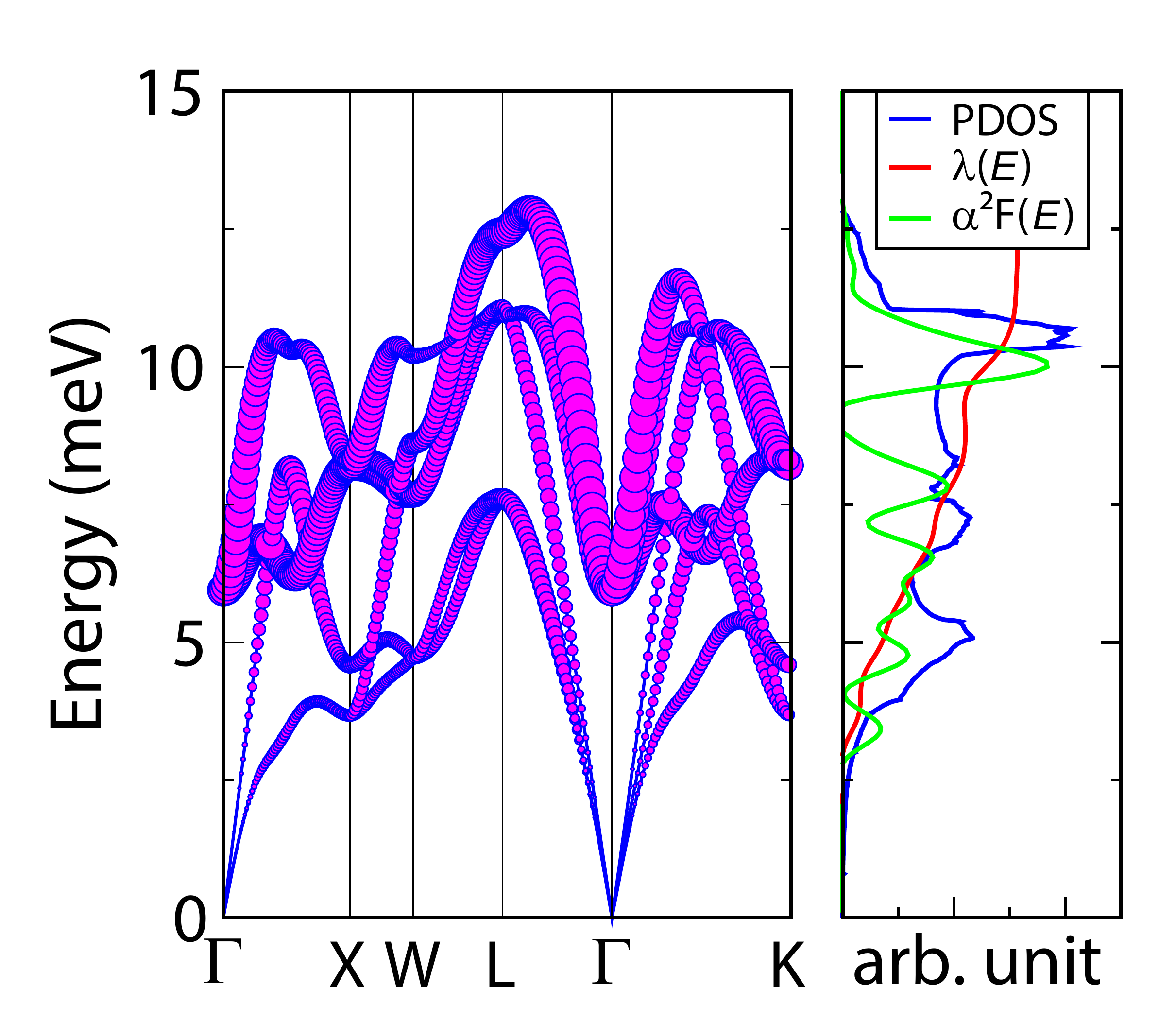}
	\caption{Left:  Phonon dispersions for \psit\ with $x=0.3$ using density functional perturbation theory and the virtual crystal approximation \cite{Ran_2018}.  Area of symbols is proportional to the calculated electron-phonon scattering rate.  Right: curves showing the calculated phonon density of states (PDOS, blue), the Eliashberg function $\alpha^2F(E)$ for $E=\hbar\omega$ (green), and the electron-phonon coupling $\lambda(E)$,  corresponding to a running integral of $2\alpha^2F(E)/E$ (red).}
	\label{fg:el-ph}
\end{figure}

The rock-salt structure of SnTe is unstable to a distortion in which the Sn sublattice is displaced along a [111] direction relative to the Te, resulting in a phase with rhombohedral symmetry, as in GeTe \cite{Pawley_1966,Oneill_2017}.  The lattice distortion lowers the electronic energy \cite{Rabe_1985} by allowing a spatially-anisotropic redistribution of the charge in the bands associated with the Sn $5s$ and $5p$ bands \cite{Waghmare_2003}; in particular, there is a maximum of the calculated Sn $5s$ charge density in the direction of the three long Sn-Te bonds, which allows enhanced hybridization of the Sn $5p$ and Te $5p$ states along the three short Sn-Te bonds.   It follows that the structural transition is electronically driven.  At the transition, the frequency of the TO mode goes to zero \cite{Oneill_2017}.   If the compound is doped away from the insulating state, one would expect the soft phonons to show a strong coupling to electrons.  Figure~\ref{fg:el-ph} shows the dispersion of the phonons indicated by circles whose area is proportional to the electron-phonon scattering rate, as calculated for \psit\ with $x=0.3$ by density functional perturbation theory, with pseudopotentials determined by the virtual crystal approximation \cite{Bellaiche_2000}, and described in \cite{Ran_2018}; similar results have been reported recently for In-doped SnTe \cite{Nomoto_2020}.  As one can see, the TO mode and especially the LO mode have a strong electron-phonon coupling in the vicinity of the $\Gamma$ point.  Moreover, the metallic state causes the TO and LO frequencies to be equal at $\Gamma$, where they are calculated to be reasonably soft.

PbTe, in contrast, is stable in the rock-salt structure; however, a recent theoretical analysis \cite{Sangiorgio_2018} supports experimental evidence \cite{Bozin_2010,Jensen_2012} for dynamic off-site Pb-Pb correlations in $\langle100\rangle$ directions.  Again, the calculations show that, within these correlated distortions, the Pb $6s$ charge density tends to pile up on the long-bond side \cite{Sangiorgio_2018}.  Now, PbTe also exhibits a reasonably soft TO mode that has anharmonic interactions with the longitudinal acoustic (LA) mode \cite{Delaire_2011,Li_2014}.  Several theoretical studies have addressed these anharmonic effects \cite{Zhang_2011,Chen_2014,Ribeiro_2018}.  The anharmonic effects are particularly relevant to achieving low thermal conductivity and high thermoelectric performance \cite{Zhao_2016}.  In our case, we are interested in superconductivity, and hence the role of electronic states driving, or coupled to, lattice instabilities is of key relevance.

\section{Experimental Details\label{ED}}

Single crystals of \psit\, $x = 0$ and $0.2$, were grown using the modified floating-zone technique discussed in Ref.~[\citenum{Zhong_2014,Zhong_2017}]. \psit\ has the cubic rocksalt structure with $Fm\bar{3}m$ crystal symmetry; room-temperature lattice parameters are $ a = b = c = 6.392(2)$ and $6.372(2)$~\AA\ for $x = 0$ and $0.2$, respectively.  We will express the momentum transfer $ \textbf{Q} = (H, K, L)$ in reciprocal lattice units (rlu) given by $(2\pi/a,2\pi/b,2\pi/c)$, 
and the momentum transfer within the first Brillouin Zone is $\textbf{q} = \textbf{Q} - \bm{\tau} = (\xi,\eta,\zeta)$, where $\bm\tau$ is a reciprocal lattice vector.

INS measurements were carried out on a non-superconducting single crystal \pst\ (mass $ = 20$~g) at the HYSPEC spectrometer at the Spallation Neutron Source (SNS) at Oak Ridge National Laboratory, and on a superconducting single crystal \psit\ $x = 0.2$ (mass $=6$ g), with $T_\mathrm{c} = 3.8$~K \cite{Zhong_2014}, at the CNCS spectrometer \cite{Ehlers_2011,Ehlers_2016} at SNS and the SPINS spectrometer at the NIST Center for Neutron Research.  Each crystal was aligned with $(H, H, 0)$ and $(0, 0, L)$ in the horizontal scattering plane; the vertical direction corresponds to $[K, -K, 0]$. Samples were cooled using a closed-cycle He cryostat at HYSPEC and an orange He-flow cryostat at both CNCS and SPINS.

At HYSPEC, measurements were carried out with incident energies of $E_{\textrm{i}}= 20$ and $27$~meV at $T=5$ and $300$~K; they involved rotating the sample about its vertical axis in $1\degree$ steps over $\approx 130/135 \degree$ range. At CNCS, the measurement was carried out with $E_{\textrm{i}}= 12$~meV at $T=1.7$~K, and by rotating the sample about its vertical axis in $1\degree$ steps over $\approx 130 \degree$ range. At SPINS, the measurements were carried out with a fixed final energy of $5$~meV and using horizontal beam collimations of Guide-80'-S-80'-open (S = sample) together with a cooled Be filter after the sample to reduce higher-order neutrons. Moreover, the measurements were done at $T=1.5, 5$, and $20$~K. Data were visualized using the \textsc{dave mslice} \cite{Dave} and \textsc{horace} \cite{Horace} software packages.

Thermal conductivity of an $x=0$ crystal was measured in a Physical Property Measurement System (PPMS, Quantum Design) using the Thermal Transport Option (TTO) from 2.1 to 300~K. Dynamic mode was used for establishing the steady state under a pre-fixed temperature gradient ($\Delta T/T = 1$-3\%).  A Ni standard provided by Quantum Design (QD) was used for thermal conductivity calibration. Small humps in the thermal conductivity versus temperature curve, noticeable at the temperatures around 75 and 275~K are related to the change of temperature gradient and to the relaxation-time fitting constant exceeding the limit of time period given by the equipment, a known artifact in Quantum Designs TTO.

\section{Results and Analysis}

\subsection{Nuclear Bragg Peaks\label{NBP}}

It is instructive to consider the conditions for Bragg scattering in the rocksalt structure.  The structure factor is given by
\begin{equation}
F(H,K,L) =
\begin{cases}
4b_A + 4b_{\rm Te},& \text{$H, K, L$ all even}\\
4b_A - 4b_{\rm Te},& \text{$H, K, L$  all odd}\\
0, & \text{otherwise}
\end{cases}       
\end{equation}
where $b_{\rm Te}$ is the nuclear scattering length for Te and $b_A$ is the appropriate average scattering length for Pb, Sn, and In.  These two distinct characters indicate that, at least for phonons near zone center, the intensities of acoustic modes (where neighboring atoms are moving in phase) will be largest when $H$, $K$, and $L$ are all even, while the intensities of optical modes (where neighboring atoms move against each other) should be largest when $H$, $K$, and $L$ are all odd.  Note that if we had ordering of the Pb and Sn atom on the $A$ sublattice, it would break the symmetry and we would have a finite structure factor for the case of mixed even and odd indices.

\begin{figure}[]
	\centering
	\includegraphics[width=0.7\columnwidth]{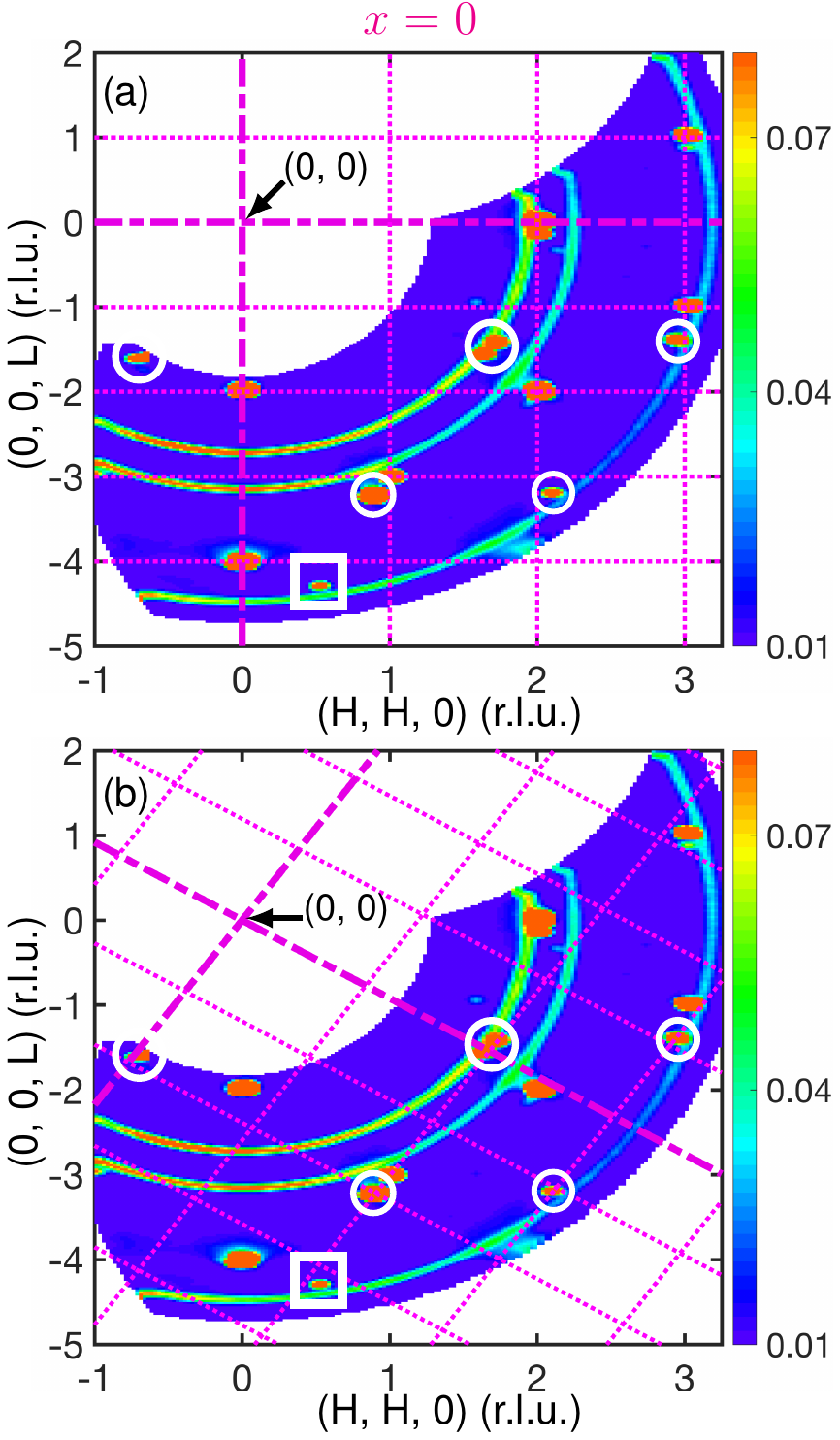}
	\caption{(a) Elastic neutron scattering spectra for \pst\ measured with an incident energy $E_\mathrm{i} = 20$~meV at $T = 5$~K. Peaks enclosed by white circle and rectangle are from the secondary grains present in the sample. (b) Same data as in (a) but with the coordinate grid rotated by $\sim 33^{\circ}$ to highlight the indexing of peaks from the secondary grain indicated by circles. The weak peak enclosed by a rectangle appears to be from a third grain. }
	\label{BP}
\end{figure}

Figure~\ref{BP}(a) shows the elastic scattering for the $x=0$ sample in the $(H,H,L)$ zone.  While there is no intensity at positions corresponding to mixed even and odd indices, indicating a random arrangement of Pb and Sn atoms, we do see another set of peaks indicated by white circles, as well as one peak indicated by a square. Figure~\ref{BP}(b) shows that the white-circle peaks correspond to a second grain that is rotated by $\sim33\degree$ with respect to the first grain. The white-rectangle peak appears to be a (3,3,1) reflection from a small third grain with a rather different orientation. While it is not optimal to have the extra, misoriented grains in the sample, it is nevertheless possible to identify, with care, the phonon modes of the main grain, as we will show. A similar situation also applies to the $x=0.2$ crystal.

\subsection{Phonon Dispersions\label{PDET}}

\subsubsection{Normal modes in \pst}

\begin{figure*}[t]
	\centering
	\includegraphics[width=1.0\textwidth]{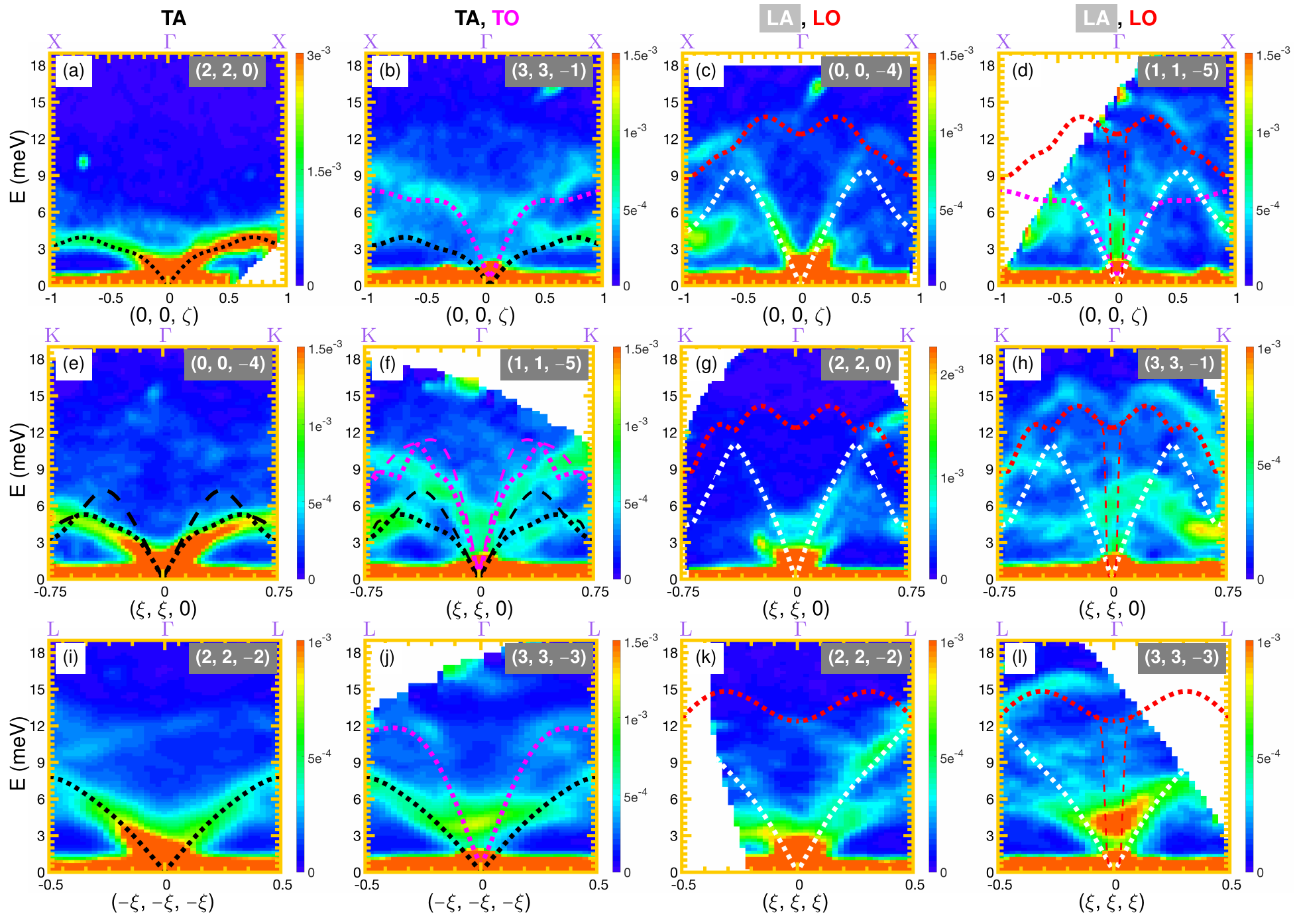}
	\caption{Color contour plots of inelastic scattering intensity for \pst, measured at $T = 5$~K with either $E_\mathrm{i} = 20$ \big[(a) and (g)\big] or $27$~meV (others), illustrating the phonon dispersions along three high-symmetry directions, $(0,0,\zeta)$, $(\xi,\xi,0)$, and $(\xi,\xi,\xi)$.  Lines display the calculated phonon dispersions \cite{Ran_2018}. Black, magenta, white and red lines correspond to transverse acoustic (TA),  transverse optical (TO), longitudinal acoustic (LA) and longitudinal optical (LO) modes. Same color dotted and dashed lines represent two branches, that are otherwise degenerate.  The Bragg point corresponding to $\Gamma$ is listed in the upper right of each panel.  Labels at the top of each column (TA, TO, etc.) highlight the dominant phonon branches in the corresponding panels.
	}
	\label{PD}
\end{figure*}

Figure~\ref{PD} presents the low-temperature inelastic data for the  \pst\ sample as color contour maps.  From top to bottom, we show phonon dispersions along $(0,0,\zeta)$, $(\xi,\xi,0)$, and $(\xi,\xi,\xi)$.  From left to right, we plot results measured transverse (or close to transverse) to Bragg points (where the scattering cross section should select transverse-polarized modes), followed by longitudinal measurements.  Data obtained at Bragg points with all even Miller indices are dominated by the acoustic modes (columns 1 and 3 of Fig.~\ref{PD}), as expected.  To see the optical modes, we have to look at the data obtained near Bragg points with indices that are all odd (columns 2 and 4 of Fig.~\ref{PD}).  For comparison, we have overplotted, as dashed lines, the relevant calculated phonon dispersions.

In the first column, we see that the signal is strongly dominated by the TA mode, as expected, and that it is in good qualitative agreement with the calculated dispersion.  The second column shows the TO mode, as well as some weight from the TA.  The calculated TO mode has a zone-center energy of just under 1 meV, whereas the data show a broad peak centered near 4 meV (we will return to this later).  Looking at the measurements of the LA mode in the third column, we see that there also appears to be significant weight from the TA mode; this is likely due to the coarse vertical resolution at HYSPEC, associated with vertical focussing of the incident neutron beam.  If we compare the results for the LO mode in column 4 with the LA results in column 3, we can see evidence from the measured intensities that there must be an avoided crossing between the LA and LO modes along the $(0,0,\zeta)$ and $(\xi,\xi,0)$ directions.  The weight that follows the calculated LA mode from zone center appears to match more closely the LO branch at the zone boundary [Fig.~\ref{PD}(c) and (g)], and the converse jump from LO to LA is apparent in the last column [Fig.~\ref{PD}(d) and (h)].

\begin{figure*}[t]
	\centering
	\includegraphics[width=0.8\textwidth]{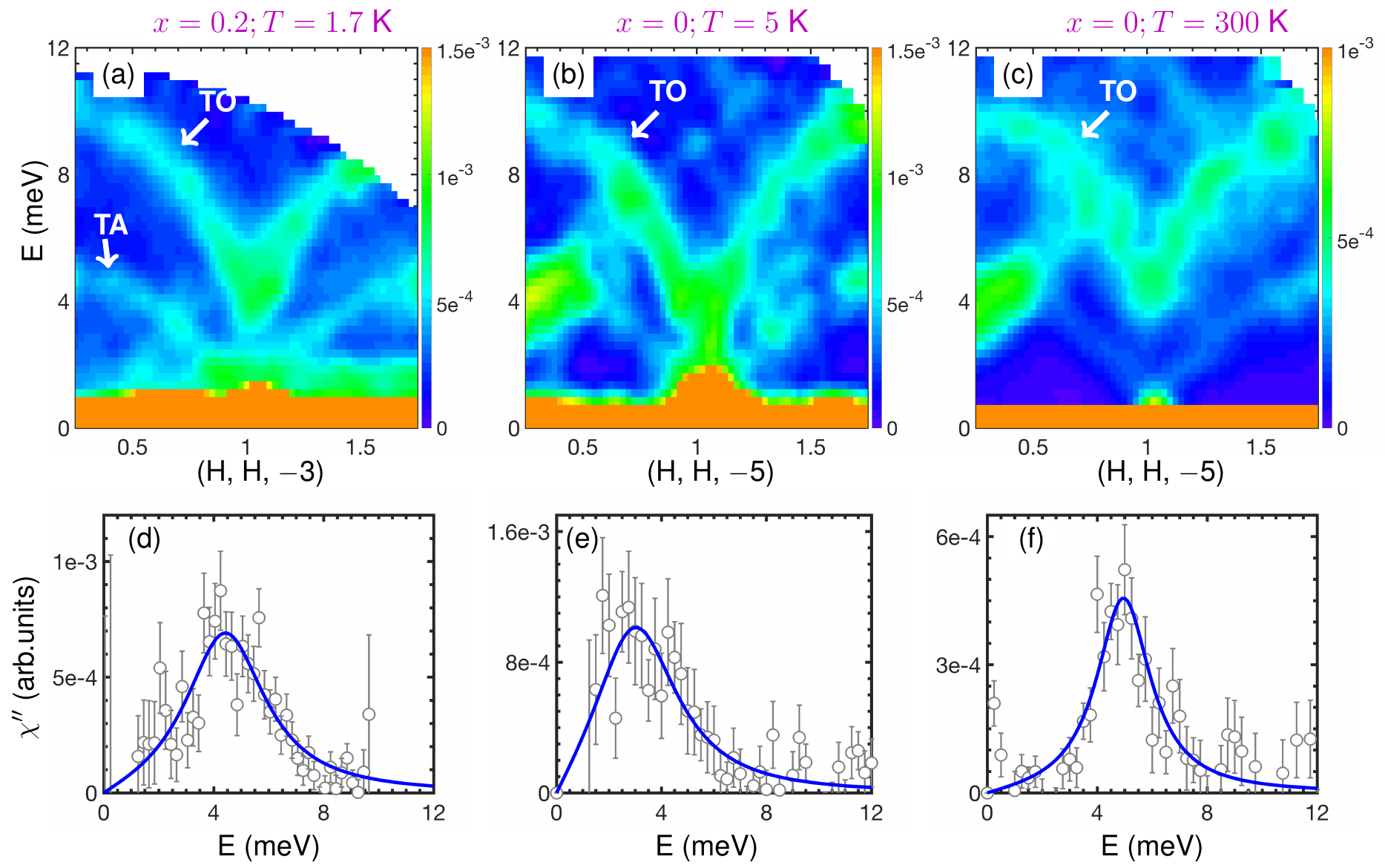}
	\caption{Color contour maps of $\chi''({\bf Q},E)$ for (a) the superconducting $x=0.2$ sample at 1.7~K (measured at CNCS with $E_i=12$~meV near ${\bf Q}=(1,1,-3)$), (b) the $x=0$ sample at 5 K, (c) and at 300~K (measured at HYSPEC with $E_i=27$~meV near $(1,1,-5)$). [The strong feature near $H=0.5$ in (b) and (c) is a TA mode from the $(2,2,-4)$ peak of the second grain.]   Corresponding constant-{\bf Q} cuts at ${\bf q}=0$ are plotted in (d), (e), and (f).  Here, the data are averaged over $\pm0.05$~rlu in each in-plane direction of {\bf Q} and $\pm0.07$ rlu out-of-plane, and a constant background, determined from the energy-gain side, has been subtracted.  Blue lines show fits to a damped-harmonic-oscillator function; fitting parameters are given in Table~\ref{TabTO}.}
	\label{wf}
\end{figure*}

As already mentioned, the model calculation was done for an insulating phase, which allows the LO energy at $\Gamma$ to be large and different from the TO.  In contrast, our sample is metallic, so we expect that the LO and TO modes must have equal energies at $\Gamma$, which would pull the LO energy far below the calculated value.  Looking at the data, we see relatively strong LO intensity for intermediate wave vectors, but it weakens on approaching $\Gamma$.  (Keep in mind that the phonon intensity is inversely proportional to the energy.)  In that vicinity, we see some distribution of intensity spread over a large energy range.  To indicate what we expect might be calculated for a metallic model, we have plotted (with a thin dashed line in the $4^\mathrm{th}$ column) a steep dispersion down to the calculated TO energy at $\Gamma$.  While the data do not show a clear continuous dispersion of this kind, we do see some extra intensity near $\Gamma$ for $E \lesssim 12$~meV.

The qualitative agreement with the calculated dispersions indicates that the observed normal modes correspond to the average lattice.  Our data also appear to be a rough interpolation between previous results reported for PbTe and SnTe \cite{Cowley_1969,Delaire_2011,Li_2014,Oneill_2017}.  

\subsubsection{Temperature and doping dependence of soft TO mode}

We compare the doping and temperature dependence of the TO mode in Fig.~\ref{wf}.  The top row shows color contour maps of the TO dispersion along a [110] direction for the superconducting $x=0.2$ sample at 1.7~K, and for the $x=0$ sample at 5~K and 300~K.  To make a quantitative comparison of the $q=0$ excitations, constant-{\bf Q} cuts, with fits to a damped harmonic oscillator, are shown in the bottom row; the resulting peak energies and widths are listed in Table~\ref{TabTO}.  The lowest TO energy occurs for the $x=0$ sample at low temperature, where it is overdamped. It hardens by 1.5~meV and sharpens on warming to room temperature.  The case of the superconducting sample at 1.7~K is somewhat intermediate with respect to those.

\begin{table}[b]
	\caption{Parameters obtained from fits of the damped-harmonic oscillator formula to the TO minima shown in Fig.~\ref{wf}(d)-(f); FWHM = full width at half maximum.  
	\label{TabTO}}.
	\begin{ruledtabular} 
	\begin{tabular}{cccc}
	sample & $T$ & $E_{\rm TO}$ & FWHM \\
	$x$ & (K) & (meV) & (meV) \\
	 \hline
	 0.2 & 1.7 & $4.8\pm0.2$ & $3.6\pm0.5$ \\
	 0.0  & 5 & $3.6\pm0.2$ & $4.0\pm0.7$ \\
	 0.0 & 300 & $5.1\pm0.2$ & $2.4\pm0.4$ \\
	\end{tabular}
	\end{ruledtabular}
\end{table}

In the case of SnTe, the TO energy at the $\Gamma$ point softens to zero at the transition to the ferroelectric state \cite{Oneill_2017}.  Based on measurements demonstrating the absence of such a displacive transition in \pbsnte\ at small $y$, the critical concentration for the completed transition was estimated to be near $y=0.5$ \cite{Daughton_1978}.   At the same time, the low-temperature $E_{\rm TO}$ of our $y=0.5$, $x=0$ crystal is slightly larger than that reported for $y=0.2$, and is actually comparable to that in PbTe \cite{Li_2014}, indicating that the trend is not trivial to estimate simply on the basis of composition.

The measurements on the $x=0$ sample in Fig.~\ref{wf}(b) and (c) are in a zone where the TO structure factor is large but the TA is small.  Hence, it is interesting to note that the $\chi''$ amplitude of the TA mode near $\Gamma$ seems to be enhanced at 5~K compared to 300~K, which might indicate some mixing of the TO character into the TA as the TO mode softens.  While part of the TA amplitude change is partially compensated by the temperature dependence of the Debye-Waller factor, comparisons of constant-{\bf Q} cuts suggest that there may be a real effect.  This may deserve attention in a future study.

As we have mentioned, our samples are essentially metallic, so we expect to have weight from the LO mode overlapping with the TO at $\Gamma$.  Of course, as we have seen for the $x=0$ sample in Fig.~\ref{PD}, the LO mode has to connect to the branch up at $\sim15$~meV just a short distance from $\Gamma$.  This may take the form of the ``waterfall'' effect seen for the TO mode in relaxor ferroelectrics \cite{Gehring_2000,Stock_2018}.  While we do see hints of weak signal near $\Gamma$ at energies between 7 and 12~meV in Fig.~\ref{PD}(d), (h), and (l), spreading a modest amount of spectral weight over such a large energy range may require more counting time to detect convincingly. 

\subsubsection{Thermal conductivity  \label{LTC}}

\begin{figure}[b]
	\centering
	\includegraphics[width=\columnwidth]{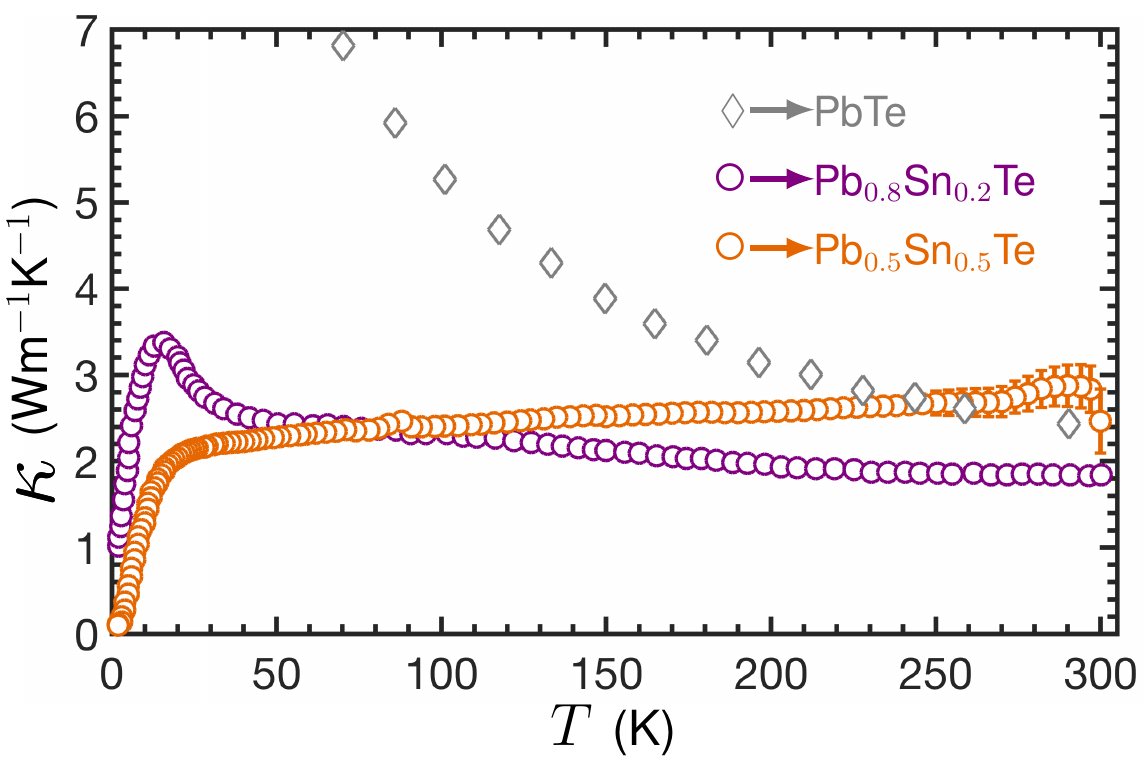}
	\caption{Temperature dependence of the thermal conductivity for \pbsnte\  with $y = 0, 0.2$ and 0.5. Orange circles: our measurement on $y=0.5$; purple circles: results for $y=0.2$  from Fig.~2 of \cite{Kim_2019}; gray diamonds: results for $y=0$ from Fig.~2 of \cite{Morelli_2008}.}
	\label{TC}
\end{figure}

Figure~\ref{TC} shows the temperature dependence of the thermal conductivity, $\kappa$, measured on our sample of \pbsnte\ with $y=0.5$, in comparison with reported results for samples with $y=0$ \cite{Morelli_2008} and 0.2 \cite{Kim_2019}.  The $\kappa(T)$ for our sample does not show a peak at low $T$, in contrast to the $y=0.2$ data, and it gradually rises with $T$, becoming comparable in magnitude to $\kappa$ for PbTe at room temperature.

How do we understand this behavior and what is its significance?  In an insulator or semiconductor, the total thermal conductivity $\kappa$ is largely determined by the phonons at low to intermediate temperatures.  Starting at very low temperature, $\kappa$ rises rapidly with $T$ as the range of thermally-excited acoustic modes increases. Eventually, scattering between phonons due to anharmonic effects limits $\kappa$, which then decreases, often as $1/T$, at higher temperatures \cite{Ashcroft}. Such a decrease is apparent in Fig.~\ref{TC} for the $y=0$ and 0.2 samples.  

In a metal, electrons typically dominate the thermal conductivity.  Our sample \pbsnte\ $y=0.5$ is metallic, but with a fairly high resistivity of approximately 1 m$\Omega$~cm \cite{Zhong_2014}. The fact that $\kappa(300~K)$ is slightly larger that that for PbTe is presumably due to the electronic contribution. The smaller magnitude of $\kappa$ at lower temperatures must be due to a small phonon contribution.

PbTe based compounds are of interest because of their potential applications in thermoelectric devices, where the figure of merit is inversely proportional $\kappa$ \cite{Zhao_2016}. As such, PbTe has an unusually low thermal conductivity \cite{Morelli_2008} and diffusivity \cite{Martelli_2018}. The analysis of phonon anomalies measured by INS led to the conclusion that the low $\kappa$ in PbTe is due to coupling between LA and TO modes near to $\Gamma$ \cite{Delaire_2011}. In that case, the energy minimum of the bare TO mode is higher than in our $y=0.5$ case; also, PbTe is insulating, so that the LO mode is even higher. Our much lower $\kappa$ suggests substantial scattering of acoustic phonons from the soft optical modes.  

Kim {\it et al.} \cite{Kim_2019} studied the Hall mobility, as well as the thermal conductivity, in \pbsnte\ with $y=0.2$ to 0.4. They observed decreasing Hall mobilities with increasing $y$, along with reduced low-temperature $\kappa$.  They emphasized the role of electron-phonon coupling, and discussed a possible polaronic interpretation of the transport.  The latter concept would have interesting implications for the superconductivity in our In-doped samples.

\subsubsection{Lowest-energy dispersive modes \label{DMLZBE}}

In the experimental phonon densities of states (PDOSs) determined for \psit, $x = 0$ and $0.3$, additional peaks were observed at energies below the lowest-energy peak in the calculated PDOS \cite{Ran_2018}.  Peaks in the density of states are typically van Hove singularities associated with well-defined modes that have a flat dispersion at the Brillouin zone boundary.  Do the new features correspond to unexpected dispersive modes at low energy?  While none are apparent in Fig.~\ref{PD}, we take a closer look here.   

\begin{figure}[t]
	\centering
	\includegraphics[width=\columnwidth]{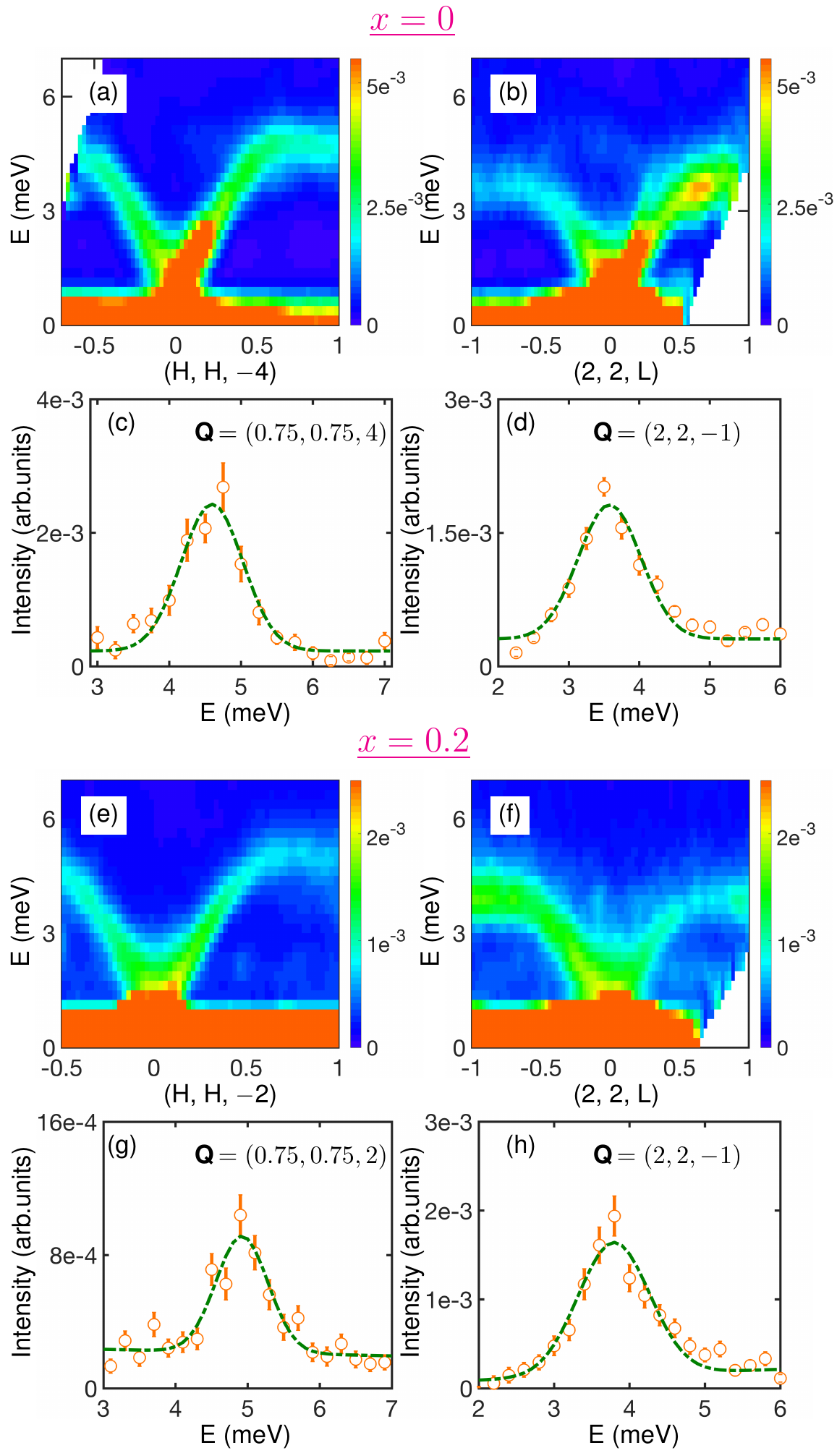}
	\caption{Transverse acoustic phonons for the  $x = 0$ (top) and $x = 0.2$ (bottom) samples.  Results for $x=0$ were done at HYSPEC with $T = 5$~K and $E_i = 20$~meV; $x=0.2$ results are from CNCS with $T = 1.7$~K and $E_i = 12$~meV at CNCS. (a), (b), (e) and (f) show the TA phonon dispersions in zones $(0, 0, -4)$, $(2, 2, 0)$, $(0, 0, -2)$ and $(2, 2, 0)$, respectively. (c), (d), (g) and (h) show constant-{\bf Q} cuts at the corresponding zone boundary points. Green dashed lines are fits to a Gaussian line shape plus background.}
	\label{TP}
\end{figure}

One can see from Fig.~\ref{PD} that the lowest-energy zone-boundary phonons (along high-symmetry directions) are the TA modes at the X and K points.  In Fig.~\ref{TP}, we present plots of the associated phonon dispersions on an expanded scale for the both the $x=0$ and $x=0.2$ samples.  The figure also includes constant-{\bf Q} cuts at the zone-boundary points, with gaussian fits indicated by dashed lines; the phonon energies from the fits at listed in Table~\ref{Tab1}.  The latter values interpolate between the values reported for PbTe and SnTe \cite{Li_2014} and are  in good agreement with the first peak energies with $E>3$~meV observed in the PDOS reported in the powder measurement \cite{Ran_2018}.  Hence, we conclude that the lower-energy features in the PDOS are not associated with sharply-defined dispersive modes.

\begin{table}[b]
	\caption{Fitted energies for TA modes at the X and K points, corresponding to $(0,0,1)$ and $(0.75,0.75,0)$, respectively, for \psit\ obtained from the gaussian fits shown in Fig.~\ref{TP}.  
	\label{Tab1}}.
	\begin{ruledtabular} 
	\begin{tabular}{lcc}
	$x$ & $\hbar\omega_{\rm TA}$(X) & $\hbar\omega_{\rm TA}$(K)\\
	\null & (meV) & (meV) \\
	 \hline
	 0 & 3.7 & 4.6 \\
	 0.2 & 3.8 & 4.9 \\
	\end{tabular}
	\end{ruledtabular}
\end{table}

\subsection{Diffuse Scattering\label{DS}}

In our measurements, we observed not only well-defined scattering but also diffuse scattering in both elastic and inelastic channel. Below are the discussions of these diffuse scatterings.

\subsubsection{Huang scattering\label{HS}}

While the absence of Bragg peaks with mixed (odd and even) Miller indices indicates that the Pb and Sn atoms are not ordered, local structural correlations are still possible.  In particular, extended x-ray absorption fine structure studies of solid solutions of alkali halides, such as (K$_{1-x}$Rb$_x$)Br \cite{Boyce_1985}, and lead chalcogenides, such as Pb(Te$_{1-x}$Se$_x$) \cite{Lebedev_1999},  have shown that the nearest-neighbor distances follow a bimodal distribution.  For each distinct pair, the average nearest-neighbor distance is in between that of a pure compound and the average separation in the solid solution.  Because the atomic positions deviate from the average structure, there should be diffuse elastic scattering near the Bragg peak positions, commonly known as Huang scattering \cite{Huang_1947}.

\begin{figure}[t]
	\centering
	\includegraphics[width=\columnwidth]{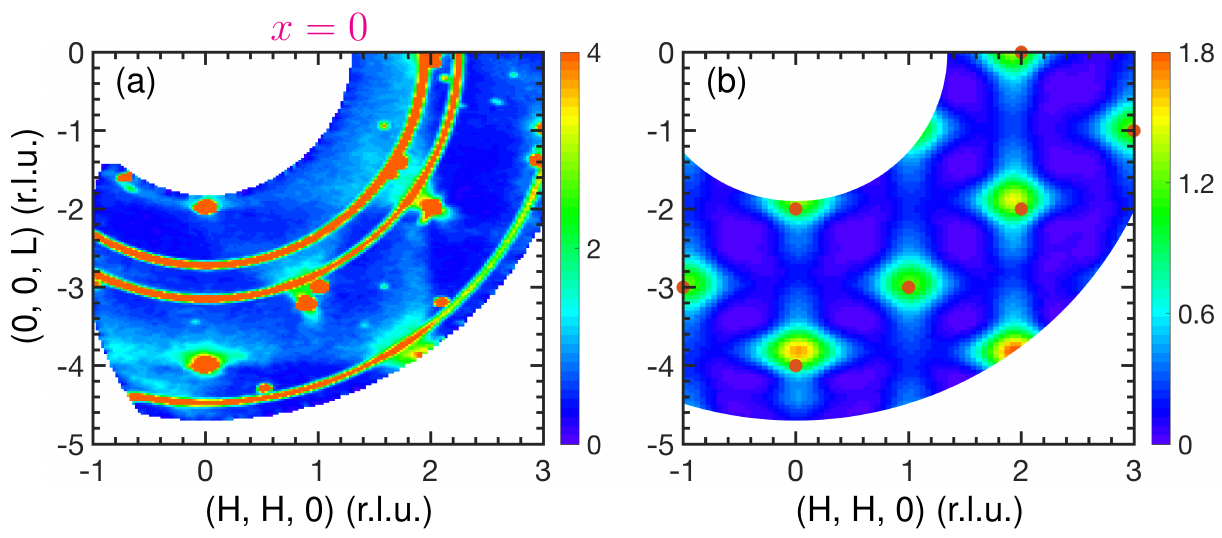}
	\caption{(a) Elastic intensity showing diffuse scattering near Bragg peaks measured on the $x=0$ sample at $5$~K.  (b) Model calculation, as discussed in the text; red dots indicate Bragg peak positions.}
	\label{hsf}
\end{figure}

\begin{figure}[b]
	\centering
	\includegraphics[width=0.5\columnwidth]{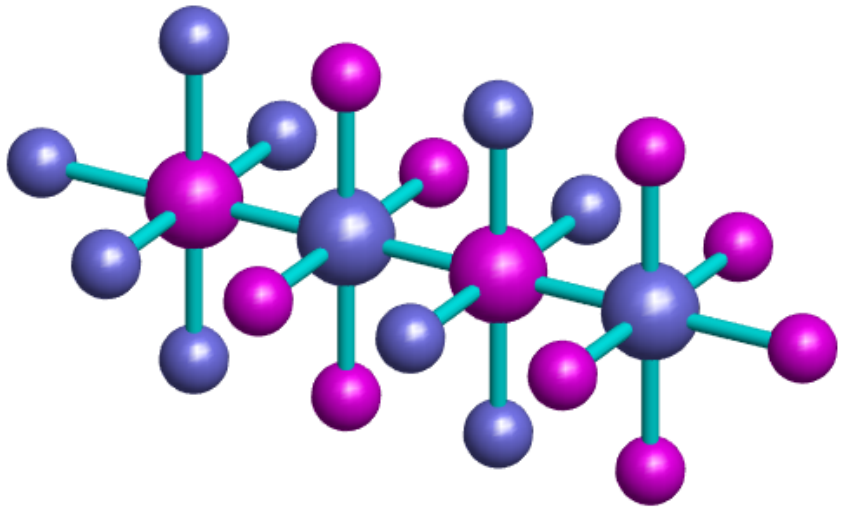}
	\caption{Illustration of the cluster used to calculate the diffuse scattering due to bond-length variations. Magenta: Sn or Pb, blue: Te.  Smaller spheres are weighted as a fraction of the large spheres.}
	\label{chain}
\end{figure}

Figure~\ref{hsf}(a) shows the elastic scattering for the $x=0$ sample at low temperature, plotted on an expanded intensity scale that allows one to see the diffuse scattering near the Bragg peaks.  Note that this scattering is rather asymmetric, appearing to the low-$Q$ side of the Bragg peaks with all-even Miller indices.  To model the scattering, we used the cluster shown in Fig.~\ref{chain}.  This represents a cluster of either SnTe or PbTe; both are included in the final result.  The central four sites have a weight of 1, while the outer 18 sites are weighted by 0.2.  The bond length equals half of the lattice parameter, which is equal to 6.466, 6.392, and 6.318~\AA\ for PbTe, Pb$_{0.5}$Sn$_{0.5}$Te, and SnTe, respectively \cite{Zhong_2013,Zhong_2014}.  To model the PbTe and SnTe bond lengths in the solid solution, we took the values of the pure materials and adjusted them to symmetrically reduce the difference by 40\%.  Starting with SnTe, we calculate the structure factor for the cluster with the appropriate bond length, and then subtract the structure factor for the average composition and bond length.  We then repeat this for PbTe, and for the different possible orientations of {\bf Q} with respect to the clusters.  The sum of the squared structure factors for each of the clusters and orientations is plotted in Fig.~\ref{hsf}(b).  

The calculation agrees qualitatively with the data.  For more quantitative agreement, one would likely need to do a calculation that does an appropriate average over all possible local clusters.  Such a calculation is beyond the scope of this study;   a qualitative understanding is judged to be sufficient for our purposes.

\subsubsection{Diffuse vibrational modes\label{LMV}}

As mentioned in the Introduction, an important motivation of the present study is to identify the nature of the unexpected low-energy features in the PDOS \cite{Ran_2018} and their possible connection with superconductivity.  We have already shown that they are not associated with standard dispersive modes.  Here we look at the inelastic diffuse scattering.

\begin{figure}[]
	\centering
	\includegraphics[width=\columnwidth]{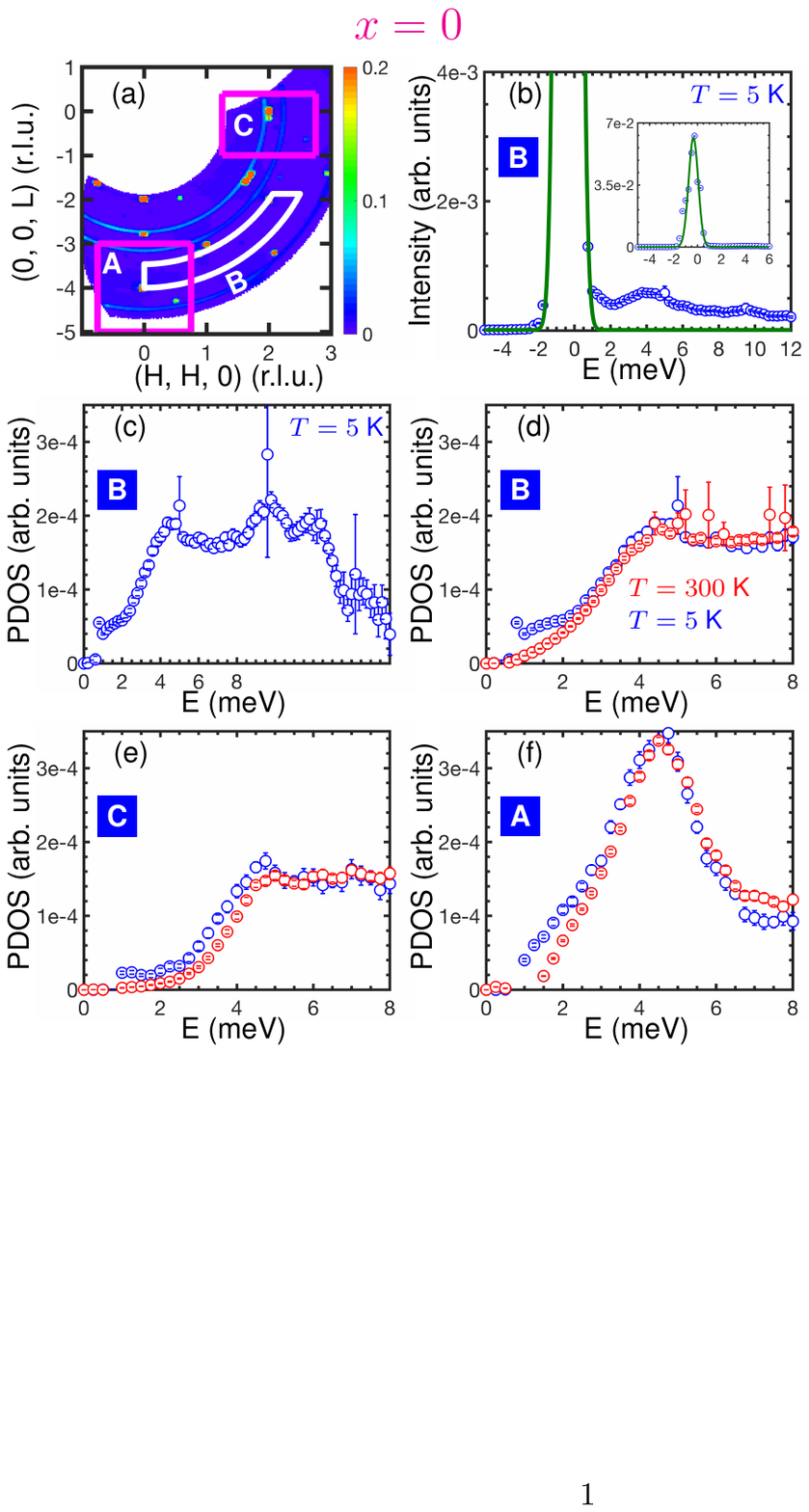}
	\caption{PDOS for non-superconducting sample $x = 0$. (a) Map of elastic scattering with boxes (labeled A, B, C) indicating regions used to evaluate the effective PDOS, as discussed in the text. (b) $S(E)$ obtained from an average over box B for $T = 5$~K; constant background estimated from the energy-gain side has been subtracted. Green solid line is a gaussian fit to the elastic peak. (c) Phonon density of states (PDOS) obtained from $S(E)$ (b) using Eq.~\eqref{Spdos}, after subtracting the fit to the elastic scattering. (d)--(f) Temperature dependence of effective PDOS evaluated for boxes B, C, and A, respectively; blue (red) symbols, $T=5$~K (300~K).}
	\label{lm}
\end{figure}

To test for the anomalous features in the single-crystal data for our $x=0$ sample, Fig.~\ref{lm}(a) shows the range of our measurements and three boxes in which we evaluated the effective PDOS $G(E)$ using the formula
\begin{equation}\label{Spdos}
\langle S(\textbf{Q}, E)\rangle \sim Q^{2}\dfrac{G(E)} {E}\big[1 + n(E)\big]
\end{equation}
where the angle brackets indicate an average over the direction of {\bf Q} and $n(E)$ is the Bose-factor.   Before extracting $G(E)$ from the angle-averaged data, a constant background determined from the energy-gain region of $-10$ to $-5$ meV for the 5-K data (where the phonon signal is negligible) was subtracted (from both the 5 and 300-K data), and the elastic scattering was fit with a gaussian and subtracted, as illustrated in Fig.~\ref{lm}(b) and (c).

The effective PDOS's obtained from the regions A, B, and C are compared for temperatures of 5 and 300 K in Fig.~\ref{lm}(f), (d), and (e), respectively.  In each case, we see a low-energy enhancement at 5~K compared to 300~K.  The energy range of extra weight is consistent with the results from powder samples \cite{Ran_2018}, but the fact that the contribution to the PDOS is reduced at 300~K is new.

\begin{figure}[]
	\centering
	\includegraphics[width=\columnwidth]{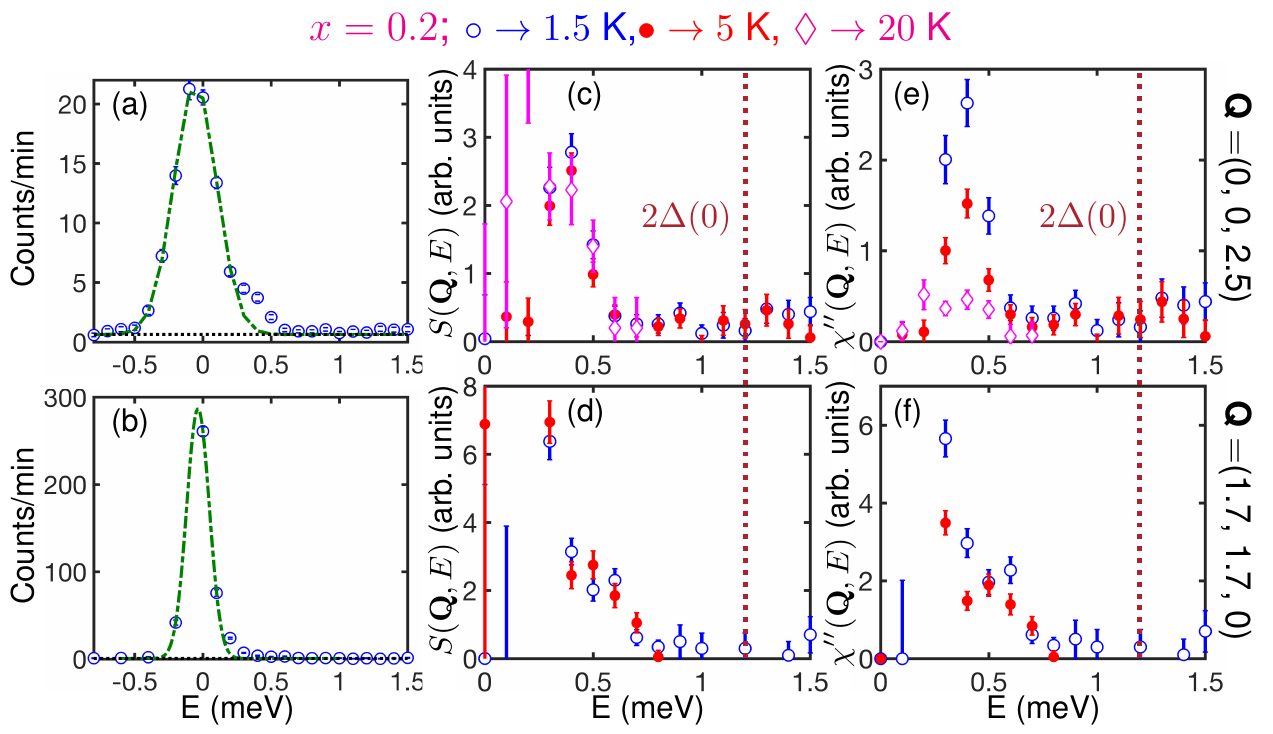}
	\caption{Inelastic scattering results on the superconducting $x = 0.2$ sample ($T_c=3.8$~K) measured at SPINS. (a),(b) Constant-\textbf{Q} scans with gaussian fits (green dashed lines) to the elastic scattering. Black dotted lines are background obtained from $-E$ (energy gain) data points. (c), (d) $S(\bm{Q}, E)$ vs. $E$ for $(0, 0, 2.5)$ and $(1.7, 1.7, 0)$, respectively, for several temperatures (legend at top). (e), (f) $\chi''({\bf Q},E)$ vs $E$ for $(0, 0, 2.5)$ and $(1.7, 1.7, 0)$, respectively. The indicated $2\Delta(0)$ (dotted maroon lines) is twice the superconducting gap at zero temperature estimated from $T_c$.}
	\label{SPINS}
\end{figure}

A striking feature in \cite{Ran_2018} is that, for a superconducting sample with $x=0.3$, the lowest-energy features in $Q$-integrated $\chi''(E)$ increase on cooling below $T_c$.  To test for such behavior in our superconducting $x=0.2$ crystal, measurements were performed on the  triple-axis spectrometer SPINS on the cold source at the NCNR.  This was not an optimal measurement for diffuse vibrational scattering because  the momentum range that could be explored was limited by the available neutron energies and only a single point can be measured at one time; nevertheless, the results shown in Fig.~\ref{SPINS} provide important clues.  

For ${\bf Q}=(0,0,2.5)$ and $(1.7,1.7,0)$, well separated from Bragg points, we observe excitations at energies below 1~meV.  When converted to $\chi''({\bf Q},E)$ (panels at right), these excitations appear to grow as the temperature goes down, but this occurs above as well as below $T_c$.  If we look instead at $S({\bf Q},E)$ (middle panels), the intensity of the excitations appears to be independent of temperature.  The weight at low energy is stronger than in the PDOS, because, as one can see from Eq.~\ref{Spdos}, $G(E)$ is corrected for the $1/E$ cross-section for phonons.  But $G(E)$ is also corrected for the thermal factor, which makes its temperature dependence similar to $\chi''({\bf Q},E)$.  If the anomalous low-energy features have an intensity that does not change with temperature, then in $G(E)$ they would appear to decrease substantially on warming to 300~K, as we observe.

\section{Discussion}

\subsection{Interpreting the diffuse excitations}

We have confirmed that the unexpected low-energy features previously discovered in the PDOS of \psit\ \cite{Ran_2018} are diffuse; there is no evidence for any new dispersive features.  These diffuse excitations are likely to be local atomic fluctuations that are made possible by atomic disorder.  They are not a trivial consequence of the disorder of Pb and Sn atoms on one sublattice of the crystal, as the collective behavior is well described by the 6 normal modes expected for the rocksalt structure.  

As already mentioned, SnTe goes through a ferroelectric transition at low temperature, with Sn atoms displaced in a [111] direction.  On warming, the lattice expands and the structure returns to the higher-symmetry cubic phase.  With Pb substitution, we expect the Sn-Te bond length to expand somewhat.  The elastic diffuse scattering provides evidence that Sn-Te and Pb-Te bond lengths are different, but studies of mixed compounds with the rocksalt structure have demonstrated that the bimodal bond lengths tend to be in between those of the pure compounds and the average of the mixed phase \cite{Boyce_1985,Lebedev_1999}.  By this argument, the Sn-Te bond length should be expanded relative to the pure compound, which would move these bonds away from structural instability.

The situation goes in the opposite direction for Pb-Te bonds.  PbTe has no structural transition; however, calculations indicate that the Pb atoms have a tendency to move off center in the [100] direction \cite{Sangiorgio_2018}.   Under pressure, PbTe transforms to an orthorhombic phase with significant displacements along [100] and a splitting of nearest-neighbor bond lengths \cite{Rousse_2005}.  Hence, the reduced Pb-Te bond lengths induced by alloying with Sn are likely to result in Pb moving to off-center positions; they enhance the electronic hybridization which can then favor modulated Pb-Te bond lengths.  Because of the lack of order on the Pb/Sn sites, collective displacements of Pb atoms are destabilized.  Thus, each Pb site may have multiple energetically-equivalent off-center sites.  It is then plausible that Pb atoms are perpetually hopping among these equivalent sites.  If there is never any ordering, then the intensity of these fluctuations will not change with temperature.  This anomalous temperature dependence has similarities to certain octahedral tilt fluctuations observed in La$_{2-x}$Sr$_x$CuO$_4$ \cite{Jacobsen_2015}. 

The low-frequency weight in the PDOS is enhanced by In doping \cite{Ran_2018}.  The In substitution results in a further reduction in average bond length \cite{Zhong_2014}, and this may increase the fraction of Pb sites that exhibit off-center fluctuations.

\subsection{Implications for superconductivity}

Our measurements show that the TO mode is fairly soft at $\Gamma$ in both our superconducting and nonsuperconducting samples.  Calculations (Fig.~\ref{fg:el-ph}) indicate that the electron-phonon coupling is significant for the TO mode, but it is even stronger for the LO mode, especially near $\Gamma$.  We have not attempted to analyze the phonon line widths here, because it is not clear that we can distinguish contributions due to electrons from effects of intrinsic elemental disorder.  Nevertheless, the small magnitude of the thermal conductivity indicates significant scattering of the acoustic modes, likely from the soft optical modes.  It is important to keep in mind that the soft TO and LO modes are due to an electronically-driven structural near instability, so that interactions that can be characterized as anharmonic are also relevant to electronic response.

Looking back at measurements of phonon line widths in the related superconductor, Sn$_{0.8}$In$_{0.2}$Te, broadening of TA phonons at small $q$ was observed \cite{Xu_2015}.  This would be consistent with coupling to the soft zone-center TO and LO modes.  To the extent that it involves electron coupling, it would suggest an interaction that involves electrons on the same small Fermi pocket, and not ``intervalley'' interactions (scattering between pockets) as originally proposed \cite{Allen_1969}.

Measurements of the resistivity for our sample compositions show that $x=0$ is more metallic \cite{Zhong_2015}.  Adding a small amount of In initially causes the compound to become extremely insulating, followed by a drop in resistivity and appearance of superconductivity; however, the superconductivity appears in the presence of a nonmetallic normal state \cite{Zhong_2017}.  The substantial resistivity could be evidence of significant electron-phonon coupling, but the lack of coherence raises questions about the applicability of the conventional model for superconductivity.  Recent studies have made the case for the ``unreasonable effectiveness'' of Eliashberg theory for non-Fermi-liquid systems \cite{Esterlis_2019B,Chowdhury_2020}.  Even in cases of low carrier density, where Migdal-Eliashberg theory breaks down, appropriate solutions to the pairing gap equations have been found \cite{Gastiasoro_2019}.  So the superconductivity in our $x=0.2$ sample appears to be covered by recent theoretical developments.

One thing that these analyses do not trivially explain, however, is why our $x=0$ sample is {\it not} superconducting.  This composition (probably doped through vacancies) is metallic and appears to have similar phonon character to the superconducting case of $x=0.2$.  The difference seems to be the presence of In, which also causes the sample to have hole-like, rather than electron-like, charge carriers.  Hirsch \cite{Hirsch_1989,Hirsch_2013} has argued that hole-like carriers are a requirement for superconductivity.  Could \psit\ be a demonstration of that?  Alternatively, does the valence-skipping character of In provide a negative-$U$ effect that drives the pairing \cite{Varma_1988}?  Or maybe another option is needed.

\section{Summary}

We have used INS to investigate the lattice vibrations in crystals of \psit, one superconducting ($x=0.2$) and one nonsuperconducting ($x=0$).  From the measurements on the $x=0$ sample, we have identified the expected set of normal modes, which are in qualitative agreement with calculated dispersions.  For both samples, we have found that the zone-center TO mode is fairly soft.  In addition, we have demonstrated that the low-energy features (below 3 meV) found previously in PDOS of powder samples correspond to weak diffuse scattering that appears below the lowest-energy TA modes at the zone boundary.  We have proposed that the inelastic diffuse scattering is associated with local vibrations of Pb atoms between off center sites.  The thermal conductivity in the $x=0$ sample is unusually low, even compared with PbTe, which indicates significant phonon scattering, which may involve interactions with both the soft optical modes and the local atomic vibrations.  

The incipient instabilities that drive the phonon anomalies are driven by electronic hybridization effects associated with unfilled $p$ states on the Sn and Pb atoms \cite{Littlewood_1979}.  These interactions are significant in both of our samples; however, only one is superconducting.  We have discussed possible  explanations of this result; however, further research is required to reach a unique conclusion.

\begin{acknowledgments}
 The work at Brookhaven was supported by the Office of Basic Energy Sciences, U.S. Department of Energy (DOE) under Contract No. DE-SC0012704. This research used resources at the Spallation Neutron Source, a DOE Office of Science User Facility operated by the Oak Ridge National Laboratory. We acknowledge the support of the National Institute of Standards and Technology, U.S. Department of Commerce, in providing the neutron research facilities used in this work. J.S.W. was supported by the National Natural Science Foundation of China with Grants No.~11822405 and 11674157, and Natural Science Foundation of Jiangsu Province with Grant No.~BK20180006. LW and JY acknowledge National Science Foundation (NSF) for its support on this project with award number 1915933.
\end{acknowledgments}

\end{document}